\documentclass[10pt]{article}
\usepackage{authblk}
\usepackage{graphicx}
\usepackage{color}
\usepackage{listings}
\usepackage{soul}
\usepackage{color}
\usepackage{bold-extra}
\usepackage{url}

\definecolor{Brown}{cmyk}{0,0.81,1,0.60}
\definecolor{OliveGreen}{cmyk}{0.64,0,0.95,0.40}
\definecolor{CadetBlue}{cmyk}{0.62,0.57,0.23,0}

\newcommand{\bad}[1]{\textcolor{red}{#1}}
\renewcommand{\bad}[1]{#1}

\newcommand{\comment}[1]{}

\newcommand{\fig}[2]{%
\begin{figure}%
\centering%
\includegraphics[width=\textwidth]{#1}%
\caption{#2}\label{#1}%
\end{figure}%
}

\newcommand{\tabref}[1]{Table~\ref{#1}}
\newcommand{\lstref}[1]{Listing~\ref{#1}}
\newcommand{\figref}[1]{Figure~\ref{#1}}
\newcommand{\secref}[1]{Section~\ref{#1}}

\newcommand{\tool}[1]{\textsc{#1}}
\newcommand{\Grammatic}[0]{\tool{Grammatic}}

\newcommand{\itemsepdelta}{%
}%
\newcommand{\subroutine}[3]
{\textsc{#1} (#2) \textbf{is}%
\begin{itemize}%
	\itemsepdelta%
	#3 %
\end{itemize}}%
\newcommand{\statement}[1]{\item #1}%
\newcommand{\var}[1]{\texttt{#1}}%
\newcommand{\foreach}[3]
{\item \textbf{for each} \var{#1} \textbf{in} \var{#2} %
\begin{itemize}%
	\itemsepdelta%
	#3%
\end{itemize}}%
\newcommand{\ifst}[3]
{\item \textbf{if} #1 %
\begin{itemize}%
	#2%
\end{itemize}%
}%
\newcommand{\call}[2]%
{\item \textbf{call} \textsc{#1}(#2)}%

\begin{document}

\lstset{
  showspaces=false,
  basicstyle=\ttfamily\footnotesize,
  keywordstyle=\bfseries\color{Brown},
  commentstyle=\color{OliveGreen},
  stringstyle=\color[rgb]{0,0,1},
  tabsize=4,
  captionpos=b,
  showstringspaces=false
}
\lstdefinelanguage{Grammatic}
	{
		morestring=[b]',
		morekeywords={lex,empty,*,?,+},
		morecomment=[l]{//},
	}

\title{Grammatical Aspects for Language Descriptions}
\author{Andrey Breslav\thanks{This work was partly done while the author was a visiting PhD student at University of Tartu, under a scholarship from European Regional Development Funds through Archimedes Foundation.}$_1$}
\affil{ITMO University$_1$\\ St. Petersburg$_1$, Russia$_1$}
\date{}

\maketitle

\begin{abstract}
	For the purposes of tool development, computer languages are usually described using context-free grammars with annotations such as semantic actions or pretty-printing instructions. 
	These descriptions are processed by generators which automatically build software, e.g., parsers, pretty-printers and editing support.

	In many cases the annotations make grammars unreadable, and when generating code for several tools supporting the same language, one usually needs to duplicate the grammar in order to provide different annotations for different generators.

	We present an approach to describing languages which improves readability of grammars and reduces the duplication. To achieve this we use Aspect-Oriented Programming principles. This approach has been implemented in an open-source tool named \Grammatic{}. We show how it can be used to generate pretty-printers and syntax highlighters.
\end{abstract}

\section{Introduction}
With the growing popularity of Domain-Specific Languages, the following types of supporting tools are created \bad{more and} more frequently:
\begin{itemize}
	\item Parsers and translators;
	\item Pretty-printers;
	\item IDE add-ons for syntax highlighting, code folding and outline views.
\end{itemize}

Nowadays these types of tools are usually developed with the help of generators which accept language descriptions in the form of annotated (context-free) grammars. 

For example, tools such as \tool{YACC} \cite{YACC} and \tool{ANTLR} \cite{ANTLR} use grammars annotated with embedded semantic actions. As an illustration of this approach first consider an annotation-free grammar for arithmetic expressions (\lstref{arithexpr}).
\begin{lstlisting}[language=Grammatic,label=arithexpr,caption=Grammar for arithmetic expressions,float]
    expr : term ((PLUS | MINUS) term)* ;
    term : factor ((MULT | DIV) factor)* ;
    factor : INT | '(' expr ')' ;
\end{lstlisting}
To generate a translator, one has to annotate the grammar rules with embedded semantic actions. \lstref{annotated_rule} shows the rule \texttt{expr} from \lstref{arithexpr} annotated for \tool{ANTLR} v3.
\begin{lstlisting}[caption=Annotated grammar rule,label=annotated_rule,float]
    expr returns [int result] : 
        t=term {result = t;} 
        ({int sign = 1;} (PLUS | MINUS {sign = -1;}) 
                               t=term {result += sign * t;})*;
\end{lstlisting}

As can be seen, the context-free grammar rule is not easily readable in \lstref{annotated_rule} because of the actions' code interfering with the grammar notation. This problem is common for annotated grammars. We will refer to it as \emph{tangled grammars}.

\fig{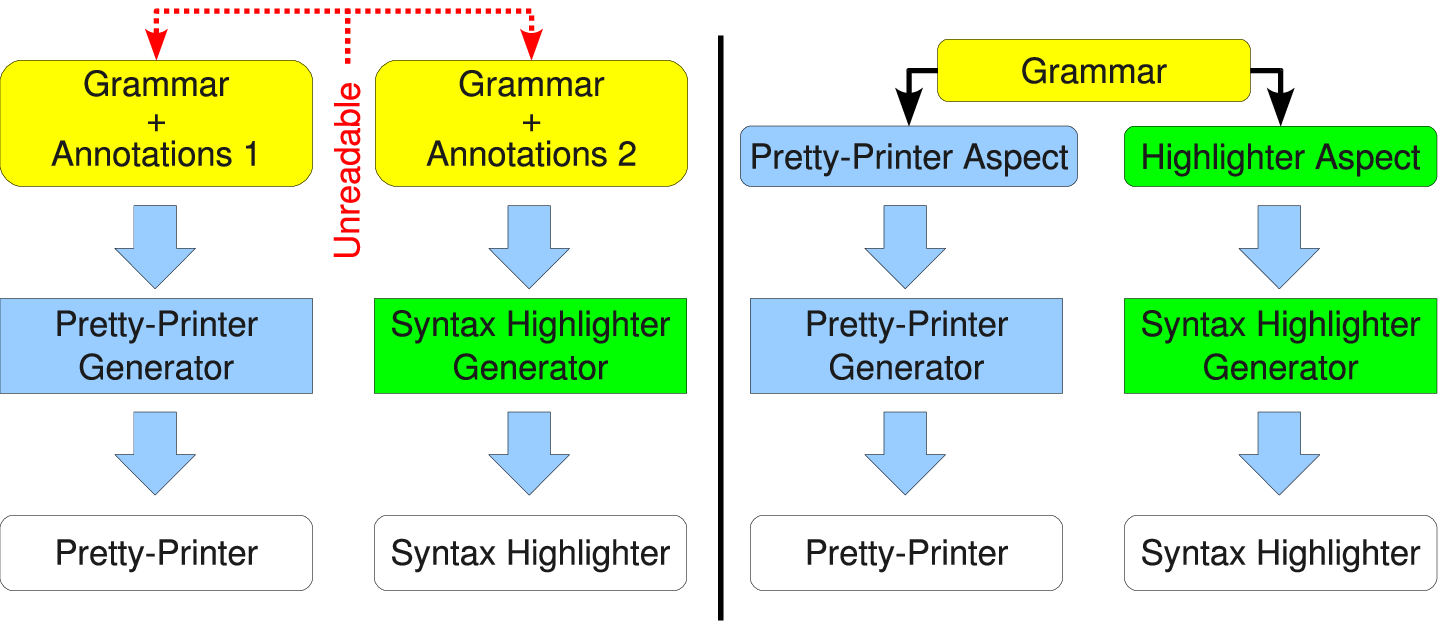}{Generating two supporting tools for the same language}
In most applications we need to create several supporting tools for the same language (see \figref{problems_and_solution}, left side). 
In such a case one uses different generators to obtain different programs (e.g., \tool{Pretzel} \cite{Pretzel} to build a pretty-printer and \tool{xText} \cite{Xtext} to create an Eclipse editor). Each generator requires its own specific set of annotations, and the developer has to write the same grammar several times with different annotations for each generator. Besides the duplication of effort, when the language evolves, this may lead to inconsistent changes in different copies of the grammar, which may cause issues which are hard to detect. We will refer to this problem as \emph{grammar duplication}.

This paper aims at reducing tangling and duplication in annotated grammars. 
A high-level view of our approach is illustrated in \figref{problems_and_solution}~(right side): the main idea is to separate the annotations from the grammar by employing the principles similar to those behind the AspectJ language \cite{AspectJ}, this leads to a notion of a \emph{grammatical aspect}. Our approach is implemented in an open-source tool named \Grammatic{}\footnote{The tool is available at \url{http://grammatic.googlecode.com}}. 

In \secref{Background} we briefly describe the main notions of aspect-oriented programming in AspectJ. 
An overview of grammatical aspects and related concepts is given in \secref{Approach}. 
\secref{Example} studies the applications of \Grammatic{} to generating syntax highlighters and pretty-printers on the basis of a common grammar. We analyze these applications and evaluate our approach in \secref{Discussion}. Related work is described in \secref{Related}. \secref{Conclusion} summarises the contribution of the paper and introduces possible directions of the future work.

\section{Background}\label{Background}
Aspect-Oriented Programming (AOP) is a body of techniques aimed at increasing modularity in general-purpose programming languages by separating cross-cutting concerns. Our approach is inspired by AspectJ \cite{AspectJ}, an aspect-oriented extension of Java. 

AspectJ allows a developer to extract the functionality that is scattered across different classes into modules called \emph{aspects}. At compilation- or run-time this functionality is \emph{weaved} back into the system. The places where code can be added are called \emph{join points}. Typical examples of join points are a method entry point, an assignment to a field, a method call. 

AspectJ uses \emph{pointcuts} --- special constructs that describe collections of join points to weave the same piece of code into many places. Pointcuts describe method and field signatures using patterns for names and types. For example, the following pointcut captures \textit{calls of all public \texttt{get-}methods in the subclasses of the class \texttt{Example} which return \texttt{int} and have no arguments}:
\begin{lstlisting}[language={[AspectJ]Java}]
    pointcut getter() : call(public int Example+.get*())
\end{lstlisting}

The code snippets attached to a pointcut are called \emph{advice}; they are weaved into every join point that matches the pointcut. For instance, the following advice writes a log record after every join point matched by the pointcut above:
\begin{lstlisting}[language={[AspectJ]Java}]
    after() : getter() {
        Log.write("A get method called");
    }
\end{lstlisting}
In this example the pointcut is designated by its name, \texttt{getter}, that follows the keyword \texttt{after} which denotes the position for the code to be weaved into.
An \emph{aspect} is basically a unit comprising a number of such pointcut-advice pairs.

\section{Overview of the approach}\label{Approach}

\Grammatic{} employs the principles of AOP in order to tackle the problems of tangling and duplication in annotated grammars. We will use the grammar from \lstref{arithexpr} and the annotated rule from \lstref{annotated_rule} to illustrate how the terms such as ``pointcut'' and ``advice'' are embodied for annotated grammars. 

\subsection*{Grammatical join points}
\figref{structured} shows a structured representation (a syntax diagram) of the annotated rule from \lstref{annotated_rule}. 
\fig{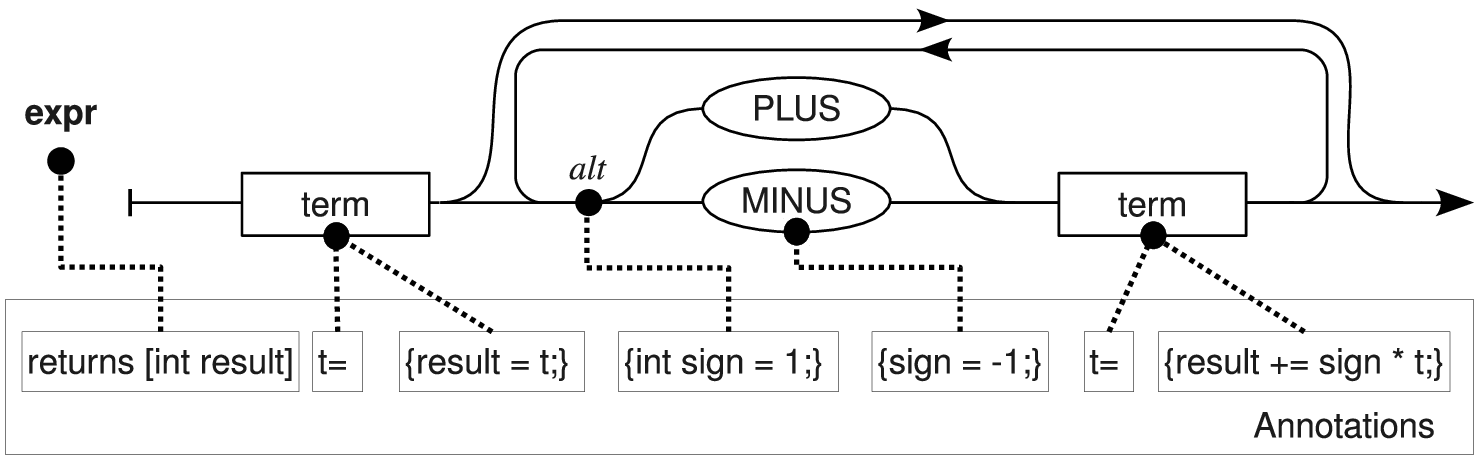}{Annotations attached to a grammar rule}
It shows the annotations attached to a symbol definition \texttt{expr}, three symbol references: \texttt{term} (two times) and \texttt{MINUS}, and an alternative~\texttt{(PLUS | MINUS)} (marked ``\emph{alt}'' in the figure). All these are examples of \emph{grammatical join points} (in \figref{structured} they are marked with black circles). The full list of join point types comprises all the types of nodes of the abstract syntax trees (ASTs) of the \emph{language of grammars}. To avoid confusion with ASTs of languages defined by the grammar, we will refer to AST of the grammar itself as \emph{grammar tree} (GT).

\Grammatic{} uses a notation for grammars which is based on the one used by \tool{ANTLR}. The only two differences are (i) in \Grammatic{} productions are explicit and separated by ``\texttt{:}'', and (ii) an empty string is denoted explicitly by ``\texttt{\#empty}''. Here is the list of types of GT nodes (which are also the types of the join points) with comments about the concrete syntax:
\begin{itemize}
	\item Grammar;
	\item Definitions of terminal and nonterminal symbols (grammar rules) and references to them;
	\item Individual productions (a rule comprises one or more productions separated by ``\texttt{:}'');
	\item Concatenation (sequence), Alternative (``\texttt{|}''), Iteration (``\texttt{*}'', ``\texttt{+}'', ``\texttt{?}'');
	\item Empty string (``\texttt{\#empty}'');
	\item Lexical literals (quoted strings).
\end{itemize}
The grammars given below may serve as example usages of this notation.

\subsection*{Grammatical pointcuts}

\Grammatic{} implements pointcuts using \emph{patterns} over the grammar language. A pattern is an expression that matches a set of nodes in GT. The syntax of the pattern language is given in \lstref{patterns}. 
\begin{lstlisting}[language=Grammatic,label=patterns,float,caption=Grammar of the pattern language]
rulePattern 
	: var? symbolPattern productionPattern* ';' ;
var 
	: '$' NAME '=' ;
symbolPattern
	: '#'        // any symbol 
	: NAME ;
productionPattern 
	: var? ':' alternativePattern
	: ':' var? '{...}' ;
alternativePattern 
	: sequencePattern ('|' (sequencePattern | (var? '...')))* ;
sequencePattern
	: iterationPattern+ ;
iterationPattern
	: var? atomicPattern ('*' | '+' | '?')? ;
atomicPattern
	: '(' alternativePattern ')'
	: symbolReferencePattern
	: '#empty'   // empty string
	: '..'       // any sequence
	: '#lex'     // any lexical literal
	: '$' NAME ; // a variable
\end{lstlisting}

The most basic form of a pattern is a direct citation from the grammar:

\lstinline!expr: term ((PLUS | MINUS) term)*;!\\
This pattern matches a rule of exactly the same form (rule \texttt{expr} from \lstref{arithexpr}). 

In addition to this capability the pattern language makes use of various types of wildcards which make patterns more abstract and thus reduce the duplication. \tabref{wildcards} summarizes available wildcards and the node types they each match.

\begin{table}[h!]
\centering
\begin{tabular}{|@{\tt}c|l|}
	\hline
	\bf Notation & \bf Matches any\ldots \\
	\hline
	\# & Symbol \\
	\#lex & Lexical literal \\
	.. & Sequence \\
	... & Nonempty set of alternatives \\
	\{...\} & Nonempty set of productions \\
	\hline
\end{tabular}
\caption{Wildcards}\label{wildcards}
\end{table}

Consider some examples of patterns for rules from \lstref{arithexpr}:
\begin{itemize}
	\item \lstinline!expr : {...}! --- a rule defining a symbol ``expr'', comprising any number of any productions (in \lstref{arithexpr} it matches only the rule for \texttt{expr});
	\item \lstinline!# : term ..! --- a production for any symbol, starting with a reference to a symbol named ``term'' (also matches only the rule for \texttt{expr});
	\item \lstinline!# : # (..)*! --- a symbol reference followed by a star iterating an arbitrary sequence (matches the rules for \texttt{expr} and \texttt{term}).
\end{itemize}

The pattern language also supports variables: a part of a pattern may be associated with a name which may be used later in the same pattern, for example:

\lstinline!# : $tr=# ((PLUS | MINUS) $tr)*!\\
Here the variable \texttt{\$tr} is defined with the pattern \texttt{\#} (any symbol) which means that all usages of the variable will match only occurrences of the same symbol. This pattern matches the rule for \texttt{expr} because the same symbol \texttt{term} is referenced in the positions matched by the variable \texttt{\$tr}. 

Note that in general a variable is bound to a \emph{set} of GT nodes: if we match the rule for \texttt{expr} against the pattern in the example above, the variable \texttt{\$tr} will be bound to a set comprised by two distinct references to the symbol \texttt{term}.

\subsection*{Grammatical advice}
Annotations attached to grammars (they are analogous to AspectJ's \emph{advice}) may have an arbitrarily complicated structure: in general, a generator may need a very rich annotation system. \Grammatic{} provides a \emph{generic annotation language}, which represents the annotations as sets of name-value pairs (see \lstref{advice}) which we call \emph{attributes}. Examples of such pairs are given in \tabref{value_types} which shows all the predefined value types. Values may also have user-defined types which can be plugged into the position marked by \texttt{<additionalValueTypes>} in the grammar.

\begin{table}[h!]
\centering
\begin{tabular}{|r@{\tt{}\,=\,}l@{\tt}|l|}
	\hline
	\multicolumn{2}{|c|}{\bf Example} & \bf Value type \\
	\hline
	int&10 & Integer \\
	str&'Hello' & String \\
	id&SomeName & Name literal\\
	rec&\{b = c; d = 5\} & Annotation\\
	seq&\{\{1, a b 'str'\}\} \,& Sequence of values \\
	\hline
\end{tabular}
\caption{Predefined value types}\label{value_types}
\end{table}

\begin{lstlisting}[language=Grammatic,label=advice,float,caption=Grammar of the advice language]
annotation
	: '{' (attribute (';' attribute?)*)? '}' 
	: '.' attribute ;
namespace
	: NAME ':' ;
attribute 
	: namespace? NAME ('=' value)? ;
value
	: character
	: INT
	: STRING
	: NAME
	: annotation 
	: '{{' (value | punctuation)* '}}'
	: <additionalValueTypes> ;
punctuation
	: '`' | '~' | '!' | '@' | '#' | '$' | '%' | '^' | '&' | '*' 
	| '(' | ')' | '-' | '+' | '=' | '|' | '\\'| '[' | ']' | ';' 
	| ':' | ',' | '.' | '/' | '?' | '<' | '>' ;
\end{lstlisting}

For example, the annotations in \figref{structured} may be represented as values of type String (other representations are also possible).

As the usage of the term ``attribute'' may be misleading in this context, we would like to note that the approach presented here does not directly correspond to attribute grammars \cite{ATG}. In fact, grammars with annotations do not have any particular execution semantics (each generator interprets the annotations in its own way), as opposed to attributed grammars which have a fixed execution semantics. One can describe attribute grammars using \Grammatic{} and define corresponding semantics in a generator, but this is just an example application.

\subsection*{Grammatical aspects}
Now, having described all the components, we can assemble a \emph{grammatical aspect} as a set of pointcuts-advice pairs.
Usage of grammatical aspects is illustrated by \figref{problems_and_solution}~(right side).

The syntax of grammatical aspects is given in \lstref{aspects}. 
\begin{lstlisting}[language=Grammatic,float,label=aspects,caption=Grammar of the aspect language]
aspect
	: grammarAnnotation? annotationRule* ;
grammarAnnotation
	: annotation ;
annotationRule
	: multiplicity? rulePattern subrules ; 
subrules
	: (subpattern | variableAnnotation)* ;
subpattern
	: '@' multiplicity? (productionPattern | alternativePattern) ':'
	                              (subrules | annotation) ;
variableAnnotation
	: '$' NAME annotation ;
multiplicity
	: '[' intOrInfinity ('..' intOrInfinity)? ']' ;
intOrInfinity
	: INT | '*' ;
\end{lstlisting}
An aspect consists of an optional \emph{grammar annotation} and zero or more \emph{annotation rules}. Annotation rules associate grammatical pointcuts (rule patterns) with advice (annotations). Here is an example of an annotation rule:
\begin{lstlisting}[language=Grammatic]
expr : $tr=# (.. $tr)*   // pointcut (pattern)
    $tr.varName = t ;    // advice (annotation)
\end{lstlisting}
In a simple case exemplified here, an annotation (\texttt{.varName = t}, the alternative syntax is \texttt{\{varName = t\}}) is attached to GT nodes to which a variable (\texttt{\$tr}) is bound.
For more complicated cases, one can define \emph{subpatterns} --- patterns which are matched against nodes situated under the matched one in the GT. For example, the following construct attaches an attribute named \texttt{varName} to each reference to the symbol \texttt{term} \emph{inside a rule matched by a top-level pattern}:
\begin{lstlisting}[language=Grammatic]
expr : ..                 // pointuct (pattern) 
    @$tr=(term):          // pointcut (subpattern)
       $tr.varName = t ;  // advice (annotation)
\end{lstlisting}
This example illustrates the typical usage of subpatters where all annotations are associated with a variable bound to the whole pattern. As a shorthand for this situation \Grammatic{} allows to omit the variable (it will be created implicitly). Using this shorthand we can abridge the previous example to the following:
\begin{lstlisting}[language=Grammatic]
expr : ..               
    @term: { varName = t } ; // '{ a = b }' is the same as '.a = b'
\end{lstlisting}
Note that subpatterns may have their own subpatterns.

Patterns and subpatterns may be preceded by a \emph{multiplicity directive}, for example
\begin{lstlisting}[language=Grammatic]
[0..1] # : $tr=# (.. $tr)* // pointcut with multiplicity
    // some advice
\end{lstlisting}
Multiplicity determines a number of times the pattern is allowed to match. The default multiplicity is \texttt{[1..*]} which means that each pattern with no explicit multiplicity is allowed to match one or more times. When an aspect is applied to a grammar, if the actual number of matches goes beyond the range allowed by a multiplicity directive, \Grammatic{} generates an error message. In the example above, such a message will be generated for the grammar from \lstref{arithexpr} because the pattern matches two rules: \texttt{expr} and \texttt{term}, which violates the specified multiplicity \texttt{[0..1]}.

\subsection*{Generation-time behaviour}

Grammatical aspects are applied at generation time. Before a generator starts working, in order to prepare the data for it, \Grammatic{} performs the following steps:
\begin{itemize}
	\statement{parse the \var{grammar} and the \var{aspect}}
	\statement{attach the \emph{grammar annotation} to the root node of the \var{grammar}}
	\foreach{annotation rule}{aspect} {
		\call{ApplyPattern}{\var{rule pattern}, \var{grammar}}
	}
\end{itemize}
Where \textsc{ApplyPattern} is a recursive subroutine defined by the following pseudocode:
\begin{center}
\parbox{0.9\textwidth}{%
\subroutine{ApplyPattern}{\var{pattern}, \var{node}}{%
	\statement{find \var{subnodes} matching \var{pattern} among descendants of \var{node}%
	\emph{\quad (Variable bindings are saved in \texttt{boundTo} map)}}%
	\ifst{the number of \var{subnodes} violates \var{pattern.multiplicity}}{%
		\statement{Report \textbf{error} and stop}%
	}{}
	\foreach{subnode}{subnodes}{%
		\foreach{subpattern}{pattern.subpatterns}{%
			\call{ApplyPattern}{\var{subpattern}, \var{subnode}}%
		}%
		\foreach{var}{pattern.variables}{%
			\foreach{boundNode}{boundTo(var)} {%
				\statement{attach \var{var.annotation} to \var{boundNode}}%
			}%
		}%
	}%
}%
\textbf{end}
}
\end{center}
The innermost loop goes through the set of GT nodes to which the variable \var{var} is bound (see \secref{Approach}) and attaches the annotations associated with this variable to each of these nodes.

If no error was reported, the resulting structure (GT nodes with attached annotations) is passed to the generator which processes it as a whole and needs no information about aspects. 

Thus, \Grammatic{} works as a front-end for generators that use its API. To use a pre-existing tool, for example, \tool{ANTLR}, with grammatical aspects, one can employ a small generator which calls \Grammatic{} to apply aspects to grammars, and produces annotated grammars in the ANTLR format.

\section{Applications}\label{Example}
In this section we show how one can make use of grammatical aspects when generating syntax highlighters and pretty-printers on the basis of the same grammar.
\subsection*{Specifying syntax highlighters}

A syntax highlighter generator creates a highlighting add-on for an IDE, such as a script for \texttt{vim} editor or a plug-in for Eclipse. For all targets the same specification language is used: we annotate a grammar with \emph{highlighting groups} which are assigned to occurrences of terminals. Each group may have its own color attributes when displayed. Common examples of highlighting groups are \emph{keyword}, \emph{number}, \emph{punctuation}. 

In many cases syntax highlighters use only lexical analysis, but it is also possible to employ light-weight parsers \cite{Island}. In such a case grammatical information is essential for a definition of the highlighter. Below we develop an aspect for the Java grammar which defines groups for keywords and for \emph{declaring occurrences} of class names and type parameters. A declaring occurrence is the first occurrence of a name in the program; all the following occurrences of that name are \emph{references}. Consider the following example:

\lstset{language=Java}
\lstinline!class!\texttt{ \ul{Example}<\ul{A}, \ul{B} }\lstinline!extends!\texttt{ A> }\lstinline!implements!\texttt{ Some<\ul{?}} \lstinline!super!\texttt{ B>} 

This illustrates how the generated syntax highlighter should work: the declaring occurrences are underlined (occurrences of \texttt{?} are always declaring) and the keywords are shown in bold. This kind of highlighting is helpful especially while developing complicated generic signatures.

\lstref{java} shows a fragment of the Java grammar \cite{JLS} which describes class declarations and type parameters. In \lstref{java_aspect} we provide a grammatical aspect which defines three highlighting groups: \emph{keyword}, \emph{classDeclaration} and \emph{typeParameterDeclaration}, for join points inside these rules.

\begin{lstlisting}[language=Grammatic,caption=Class declaration syntax in Java 5,label=java,float]
normalClassDeclaration
	: 'class' IDENTIFIER typeParameters? 
                  ('extends' type)? ('implements' typeList)? classBody ;
classBody
	: '{' classBodyDeclaration* '}' ;
typeParameters
	: '<' typeParameter (',' typeParameter)* '>' ;
typeParameter
	: IDENTIFIER ('extends' bound)? ;
bound
	: type ('&' type)* ;
type
	: IDENTIFIER typeArguments? ('.' IDENTIFIER typeArguments?)* ('[' ']')*
	: basicType ;
typeArguments
	: '<' typeArgument (',' typeArgument)* '>' ;
typeArgument
	: type
	: '?' (('extends' | 'super') type)? ;
\end{lstlisting}

\lstset{language=Grammatic}
Each annotation rule from \lstref{java_aspect} contains two subpatterns. The first one is \texttt{\#lex}: it matches every lexical literal. For example, for the first rule it matches \texttt{'class'}, \texttt{'extends'} and \texttt{'implements'}; the highlighting group \texttt{keyword} is assigned to all these literals.

The second subpattern in each annotation rule is used to set a corresponding highlighting group for a declaring occurrence: for classes and type parameters it matches \texttt{IDENTIFIER} and for wildcards --- the \texttt{'?'} literal.

When the aspect is applied to the grammar, \Grammatic{} attaches the \texttt{group} attribute to the GT nodes matched by the patterns in the aspect. The obtained annotated grammar is processed by a generator which produces code for a highlighter.

\begin{lstlisting}[language=Grammatic,caption=Highlighting aspect for class declarations in Java,float,label=java_aspect]
# : 'class' IDENTIFIER ..
	@#lex: { group = keyword } ;
	@IDENTIFIER: { group = classDeclaration } ;
typeParameter : IDENTIFIER ..
	@#lex: { group = keyword } ;
	@IDENTIFIER: { group = typeParameterDeclaration } ;
typeArgument : {...}
	@#lex: { group = keyword } ;
	@'?': { group = typeParameterDeclaration } ;
\end{lstlisting}

\subsection*{Specifying pretty-printers}

By applying a different aspect to the same grammar (\lstref{java}), one can specify a pretty-printer for Java. A pretty-printer generator relies on annotations describing how tokens should be aligned by inserting whitespace between them. 

In \lstref{pp_aspect} these annotations are given in the form of attributes \texttt{before} and \texttt{after}, which specify whitespace to be inserted into corresponding positions. Values of the attributes are \emph{sequences} (\texttt{\{\{} \ldots \texttt{\}\}}) of strings and name literals \texttt{increaseIndent} and \texttt{decreaseIndent} which control the current level of indentation. 

The most widely used values of \texttt{before} and \texttt{after} are specified in a \emph{grammar annotation} by attributes \texttt{defaultBefore} and \texttt{defaultAfter} respectively, and not specified for each token individually. In \lstref{pp_aspect} the default formatting puts nothing before each token and a space --- after each token; it applies whenever no value was set explicitly.

\begin{lstlisting}[language=Grammatic,caption=Pretty-printing aspect for class declarations in Java,label=pp_aspect,float]
{ // Grammar annotation
	defaultAfter = {{ ' ' }}; 
	defaultBefore = {{ '' }};
} 
	
classBody : '{' classBodyDeclaration* '}'
	@'{': { after = {{ '\n' increaseIndent }} } ;
	@classBodyDeclaration: { after = {{ '\n' }} } ;
	@'}': {
		before = {{ decreaseIndent '\n' }};
		after = {{ '\n' }};
	};
typeParameters : '<' typeParameter (',' typeParameter)* '>' 
	@'<': { after = {{ '' }} } ;
	@typeParameter: { after = {{ '' }} } ;
\end{lstlisting}

\section{Discussion}\label{Discussion}

This paper aims at coping with two problems: tangled grammars and grammar duplication. 
When using \Grammatic{}, a single annotated grammar is replaced by a \emph{pure} context-free grammar and a set of grammatical aspects. This means that the problem of \emph{tangled grammars} is successfully addressed.

This also means that the grammar is written down only once even when several aspects are applied (see the previous section). But if we look at the aspects, we see that the patterns carry on some extracts from the grammar thus it is not so obvious whether our approach helps against the problem of \emph{duplication} or not. Let us examine this in more details using the examples from the previous section.

From the perspective of grammar duplication, the worst case is an aspect where all the patterns are exact citations from the grammar (no wildcards are used, see \lstref{pp_aspect}). This means that a large part of the grammar is completely duplicated by those patterns. But if we compare this with the case of conventional annotated grammars, there still is at least one advantage of using \Grammatic{}. Consider the scenario when the grammar has to be changed. In case of conventional annotated grammars, the same changes must be performed once for each instance of the grammar and there is a risk of inconsistent changes which are not reported to the user. In \Grammatic{}, on the other hand, a developer can control this using \emph{multiplicities}: for example, check if the patterns do not match anything in the grammar and report it (since the default multiplicities require each pattern to match at least once, this will be done automatically). Thus, even in the worst case, grammatical aspects make development less error-prone.

Using wildcards and subpatterns as it is done in \lstref{java_aspect} (i) reduces the duplication and (ii) makes a good chance that the patterns will not need to be changed when the grammar changes. For example, consider the first annotation rule from \lstref{java_aspect}: this rule works properly for both Java version~1.4 and version~5 (see \lstref{java_14} and \lstref{java} respectively). The pointcut used in this rule is sustainable against renaming the symbol on the left-hand side (\texttt{classDeclaration} was renamed to \texttt{normalClassDeclaration}) and structural changes to the right-hand side (type parameters were introduced in Java~5). The only requirement is that the definition should start with the \texttt{'class'} keyword followed by the \texttt{IDENTIFIER}.

\begin{lstlisting}[language=Grammatic,caption=Class declaration rule in Java 1.4,label=java_14,float]
classDeclaration
	: 'class' IDENTIFIER ('extends' type)? 
                         ('implements' typeList)? classBody ;
\end{lstlisting}

In AOP, the duplication of effort needed to modify pointcuts when the main program changes is referred to as the \emph{fragile pointcut problem} \cite{Fragile}. Wildcards and subpatterns make pointcuts more \emph{abstract}, in other words, they widen the range of join points matched by the pointcuts. From this point of view, wildcards help to abstract over the contents of the rule, and subpatterns --- over the positions of particular elements within the rule. The more abstract a pointcut is, the less duplication it presents and the less fragile it is. 

The most abstract pointcut does not introduce any duplication and is not fragile at all. Unfortunately, it is also of no use, since it matches any possible join point. This means that eliminating the duplication completely from patterns is not technically possible. Fortunately, we do not want this: if no information about a grammar is present in an aspect, this makes it much less readable because the reader has no clue about how the annotations are connected to the grammar. Thus, there is a trade-off between the readability and duplication in grammatical aspects and a developer should keep pointcuts as abstract as it is possible without \bad{damaging} readability.

To summarize, our approach allows one to keep a context-free grammar completely clean by moving annotations to aspects and to avoid any unnecessary duplication by using abstract pointcuts. 

\section{Related work}\label{Related}

Several attribute grammar (AG) systems, namely \tool{JastAdd} \cite{JastAdd}, \tool{Silver} \cite{Silver} and \tool{LISA} \cite{LISA}, successfully use aspects to attach attribute evaluation productions to context-free grammar rules.
AGs are a generic language for specifying computations on ASTs. They are well-suited for tasks such as specifying translators in general, which require a lot of expressive power. But the existence of more problem-oriented tools such as \tool{Pretzel} \cite{Pretzel} suggests that the generic formalism of AGs may not be the perfect tool for problems like generating pretty-printers.
In fact, to specify a pretty-printer with AGs one has to produce a lot of boilerplate code for converting an AST into a string in concrete syntax.
As we have shown in \secref{Example}, \Grammatic{} facilitates creation of such problem-oriented tools providing the syntactical means (grammatical aspects) to avoid tangled grammars and unnecessary duplication.

The \tool{MPS} \cite{MPS} project (which lies outside the domain of textual languages since the editors in \tool{MPS} work directly on ASTs) implements the approach which is very close to ours. It uses aspects attached to the \emph{concept language} (which describes abstract syntax of \tool{MPS} languages) to provide input data to generators. The ideas behind aspects in \tool{MPS} are very close to those behind \Grammatic{} but the implementation is very different: \tool{MPS} does not use pointcuts and performs all the checking while the aspects are created.

There is another approach to the problems we address: parser generators such as \tool{SableCC} \cite{SableCC} and \tool{ANTLR} \cite{ANTLR} can work on annotation-free grammars and produce parsers that build ASTs automatically. In this way the problems induced by using annotations are avoided. The disadvantage of this approach is that the ASTs must be processed manually in a general-purpose programming language, which makes the development process less formal and thus more error-prone.

%
%
%
%
\section{Conclusion}\label{Conclusion}

Annotated grammars are widely used to specify inputs for various generators which produce language support tools.
In this paper we have addressed the problems of tangling and duplication in annotated grammars. Both problems affect maintainability of the grammars: tangled grammars take more effort to understand, and duplication, besides the need to make every change twice as the language evolves, may lead to inconsistent changes in different copies of the same grammar.

We have introduced \emph{grammatical aspects} and showed how they may be used to cope with these problems by separating context-free grammars from annotations. 

The primary contribution of this paper is a tool named \Grammatic{} which implements an aspect-oriented approach to specification of annotated grammars. \Grammatic{} provides languages for specifying grammatical pointcuts, advice and aspects.

We have demonstrated how \Grammatic{} may be used to generate a syntax highlighter and a pretty-printer by applying two different aspects to the same grammar. We have shown that the problem of tangled grammars is completely solved and all the unnecessary duplication can be eliminated. The possible negative impact of remaining duplication (necessary to keep the aspects readable) can be addressed in two ways:
	(i) \emph{abstract patterns} reduce the amount of changes in aspects per change in the grammar, and
	(ii) \emph{multiplicities} help to detect inconsistencies at generation time.

One possible way to continue this work is to support grammar adaptation techniques \cite{Laemmel} in \Grammatic{} to facilitate rephrasing of syntax definitions (e.g.,~left factoring or encoding priorities of binary operations) to satisfy requirements of particular parsing algorithms.

Another possible direction is to generalize the presented approach to support not only grammars, but also other types of declarative languages used as inputs for generators, such as UML or XSD.

\bibliographystyle{plain}
\bibliography{grammatic_ldta_10}

\end{document}